\documentclass[prd,a4paper,twocolumn]{revtex4}
\usepackage{graphicx}
\usepackage[]{subfigure}
\usepackage{epsfig}
\usepackage{graphicx}

\newcommand{\nc}{\newcommand}

\nc{\be}[1]{\begin{equation}\mbox{$\label{#1}$}}
\nc{\bea}[1]{\begin{eqnarray} \mbox{$\label{#1}$}}
\nc{\Section}[2]{\section{#2}\label{#1}}
\nc{\Bibitem}[1]{\bibitem{#1}}
\nc{\Label}[1]{\label{#1}}

\nc{\eea}{\end{eqnarray}}
\nc{\ee}{\end{equation}}

\nc{\bdm}{\begin{displaymath}}
\nc{\edm}{\end{displaymath}}
\nc{\dpsty}{\displaystyle}
\nc{\bc}{\begin{center}}
\nc{\ec}{\end{center}}
\nc{\ba}{\begin{array}}
\nc{\ea}{\end{array}}
\nc{\bab}{\begin{abstract}}
\nc{\eab}{\end{abstract}}
\nc{\btab}{\begin{tabular}}
\nc{\etab}{\end{tabular}}
\nc{\bit}{\begin{itemize}}
\nc{\eit}{\end{itemize}}
\nc{\ben}{\begin{enumerate}}
\nc{\een}{\end{enumerate}}
\nc{\bfig}{\begin{figure}}
\nc{\efig}{\end{figure}}

\nc{\arreq}{&\!=\!&}
\nc{\arrmi}{&\!-\!&}
\nc{\arrpl}{&\!+\!&}
\nc{\arrap}{&\!\!\!\approx\!\!\!&}
\nc{\non}{\nonumber}
\nc{\align}{\!\!\!\!\!\!\!\!&&}

\def\lsim{\; \raise0.3ex\hbox{$<$\kern-0.75em
      \raise-1.1ex\hbox{$\sim$}}\; }
\def\gsim{\; \raise0.3ex\hbox{$>$\kern-0.75em
      \raise-1.1ex\hbox{$\sim$}}\; }

\nc{\DOT}{\hspace{-0.08in}{\bf .}\hspace{0.1in}}
\nc{\Laada}{\hbox {$\sqcap$ \kern -1em $\sqcup$}}
\nc\loota{{\scriptstyle\sqcap\kern-0.55em\hbox{$\scriptstyle\sqcup$}}}
\nc\Loota{{\sqcap\kern-0.65em\hbox{$\sqcup$}}}
\nc\laada{\Loota}
\nc{\qed}{\hskip 3em \hbox{\BOX} \vskip 2ex}

\nc{\real}{{\rm I \! R}}
\nc{\Z}{{\sf Z \!\!\! Z}}
\nc{\complex}{{\rm C\!\!\! {\sf I}\,\,}}
\def\bigid{\leavevmode\hbox{\small1\kern-3.8pt\normalsize1}}
\def\id{\leavevmode\hbox{\small1\kern-3.3pt\normalsize1}}
\nc{\slask}{\!\!\!/}
\nc{\bis}{{\prime\prime}}
\nc{\pa}{\partial}
\nc{\na}{\nabla}
\nc{\ra}{\rangle}
\nc{\la}{\langle}
\nc{\goto}{\rightarrow}
\nc{\swap}{\leftrightarrow}

\nc{\EE}[1]{ \mbox{$\cdot10^{#1}$} }
\nc{\abs}[1]{\left|#1\right|}
\nc{\at}[2]{\left.#1\right|_{#2}}
\nc{\norm}[1]{\|#1\|}
\nc{\abscut}[2]{\Abs{#1}_{\scriptscriptstyle#2}}
\nc{\vek}[1]{{\rm\bf #1}}
\nc{\integral}[2]{\int\limits_{#1}^{#2}}
\nc{\inv}[1]{\frac{1}{#1}}
\nc{\dd}[2]{{{\partial #1}\over{\partial #2}}}
\nc{\ddd}[2]{{{{\partial}^2 #1}\over{\partial {#2}^2}}}
\nc{\dddd}[3]{{{{\partial}^2 #1}\over
    {\partial #2 \partial #3}}}
\nc{\dder}[2]{{{d #1}\over{d #2}}}
\nc{\ddder}[2]{{{d^2 #1}\over{d {#2}^2}}}
\nc{\dddder}[3]{{d^2 #1}\over
    {d #2 d #3}}
\nc{\dx}[1]{d\,^{#1}x}
\nc{\dy}[1]{d\,^{#1}y}
\nc{\dz}[1]{d\,^{#1}z}
\nc{\dl}[1]{\frac{d\,^{#1}l}{(2\pi)^{#1}}}
\nc{\dk}[1]{\frac{d\,^{#1}k}{(2\pi)^{#1}}}
\nc{\dq}[1]{\frac{d\,^{#1}q}{(2\pi)^{#1}}}

\nc{\bfT}{{\bf T }}

\nc{\cA}{{\cal A}}
\nc{\cB}{{\cal B}}
\nc{\cD}{{\cal D}}
\nc{\cE}{{\cal E}}
\nc{\cG}{{\cal G}}
\nc{\cH}{{\cal H}}
\nc{\cL}{{\cal L}}
\nc{\cO}{{\cal O}}
\nc{\cT}{{\cal T}}
\nc{\cN}{{\cal N}}
\nc{\cR}{{\cal R}}

\nc{\rvac}[1]{|{\cal O}#1\rangle}
\nc{\lvac}[1]{\langle{\cal O}#1|}
\nc{\rvacb}[1]{|{\cal O}_\beta #1\rangle}
\nc{\lvacb}[1]{\langle{\cal O}_\beta #1 |}
\nc{\bb}{\bar{\beta}}
\nc{\bt}{\tilde{\beta}}
\nc{\ctH}{\tilde{\cal H}}
\nc{\chH}{\hat{\cal H}}
\nc{\1}{\aa}
\nc{\2}{\"{a}}
\nc{\3}{\"{o}}
\nc{\4}{\AA}
\nc{\5}{\"{A}}
\nc{\6}{\"{O}}
\nc{\al}{\alpha}
\nc{\g}{\gamma}
\nc{\Del}{\Delta}
\nc{\e}{\textrm{e}}
\nc{\eps}{\epsilon}
\nc{\lam}{\lambda}
\nc{\Om}{\Omega}
\nc{\ve}{\varepsilon}
\nc{\mn}{{\mu\nu}}
\nc{\vp}{\varphi}

\nc{\B}{\mathcal B}
\nc{\ud}{\textrm{d}}
\nc{\fsky}{f_\textrm{sky}}
\nc{\fscan}{f_\textrm{scan}}

\begin{document}

\title{Constraining primordial magnetic fields with CMB polarization experiments}
\author{Jostein R. Kristiansen}
\affiliation{Institute of Theoretical Astrophysics, University of Oslo, Box 1029, 0315 Oslo, NORWAY}
\author{Pedro G. Ferreira}
\affiliation{Astrophysics, University of Oxford, Denys Wilkinson Building, Keble Road, Oxford OX1 3RH, UK}
\date{\today}

\begin{abstract}

We calculate the effect that a primordial homogeneous magnetic field, $\B_0$, will have on the different CMB power spectra due to Faraday rotation. Concentrating on the $TB$, $EB$ and  $BB$ correlations, we forecast the ability for future CMB polarization experiments to constrain $\B_0$. Our results depend on how well the foregrounds can be subtracted from the CMB maps, but we find a predicted error between $\sigma_{\B_0} = 4 \times 10^{-11}$Gauss (for the QUIET experiment with foregrounds perfectly subtracted) and $3 \times 10^{-10}$Gauss (with the Clover experiment with no foreground subtraction). These constraints are two orders of magnitudes better than the present limits on $\B_0$. 
\end{abstract}

\maketitle

\section{Introduction}   

Magnetic fields, with amplitudes of around 1$\mu$Gauss have been
observed in galaxies and clusters\cite{wolfe:1992,clarke:2000,widrow:2002,xu:2005, kronberg:2007}. Indeed it has been argued that these
fields may play a role in galaxy formation and evolution. It
is believed that these fields may have arisen from a common mechanism:
an amplification of a weak primordial field through adiabatic collapse
or some form of cosmic dynamo. There is a suite of proposals for the
origin of the primordial field, i.e. the seed field, from phase transitions, to parity
violating processes and inflation. In general the seed fields generated
by these mechanisms will be
stochastic and spatially varying. There is, however, the  possibility
that the primordial magnetic field is homogeneous, embedded in the large
scale fabric of the universe\cite{thorne:1967,turner:1987, ratra:1991,barrow:1997,Bamba:2003,bamba:2006,kunze:2007}. All of these possibilities should
lead to distinct observational signatures. In this paper we will
focus on one such signature, the polarization of the Cosmic Microwave Background
(CMB).

The polarization of the CMB is generated by Thomson scattering of 
the quadrupole temperature anisotropy at the last scattering surface. 
Since the dipole anisotropy does not generate any polarization, the 
amplitude of the polarization signal will be significantly lower than for 
the temperature anisotropies, which get most of the power from the scattering of the dipole anisotropy. This low amplitude leads to experimental challenges.
However, the different polarization power spectra will provide important and complimentary information to what we can learn from studying the temperature fluctuations alone, and up and coming experiments, such as Clover \cite{taylor:2006, north:2007}, EBEX \cite{ebex:2008,oxley:2005}, QUIET \cite{quiet:2008, samtleben:2008}, Spider \cite{mactavish:2007} and Planck \cite{planckweb:2008, planck:2005}, will attempt to do measure the CMB polarization signal with unprecedented precision.

Primordial magnetic fields leads to Faraday rotation of CMB polarization. 
This alters the overall pattern of the CMB polarization by mixing $E$ and $B$ modes. The effect can play a role on
small scales (in the case of stochastic magnetic fields)\cite{caprini:2003,durrer:2003, kosowsky:2004, lewis:2004, yamazaki:2008}
 and large scales (in the case of homogeneous magnetic fields) \cite{kosowsky:1996, harari:1996, scannapieco:1997, scoccola:2004}. In this paper we
explore the latter and forecast how well we can use future CMB polarization
experiments to constrain the amplitude of the magnetic field. 

\section{CMB power spectra in the presence of a homogeneous magnetic field}
\label{sec:cls}
The polarization of the CMB can be described in terms of the Stokes parameters, $Q$ and $U$. These parameters quantify the intensity of the incoming radiation in mutually orthogonal directions. The $Q$ and $U$ parameters are not rotationally invariant and depend on the specific choice of coordinate system. It is therefore convenient to transform to rotationally invariant $E$ (curl free) and $B$ (divergence free) modes before comparing observations to theory. These are defined by
\be
{}(Q \pm iU) (\hat n) = - \sum_{lm} (E_{lm} \pm iB_{lm}) _{\pm2}Y_{lm}(\hat n),
\ee
where $_{\pm 2}Y_{lm}(\hat n)$ denotes the spin-weighted spherical harmonics. 

The Thomson scattering process will only generate $E$ modes from scalar
perturbations of the metric and stress energy tensor. 
In the standard cosmological picture today the $B$ modes are only generated by relic gravitational waves from inflation or from gravitational lensing of $E$ modes. These two processes will induce a frequency independent $B$ mode signal for low $l$s and high $l$s respectively. $B$ modes created by gravitational waves and lensing will  lead to a non-zero $BB$ autocorrelation signal. Non-zero correlations between the $B$ mode and the $E$ or $T$ (temperature) modes can only occur under processes that break parity invariance. Thus, any detection of a non-zero $TB$ or $EB$ signal would be a sign of physics beyond the cosmological standard paradigm. 

A large-scale homogeneous magnetic field is an example of a parity violating field, and the Faraday rotation of CMB polarization induced by such a field will indeed result in non-zero $EB$ and $TB$ correlations. In an otherwise standard cosmological scenario such a magnetic field will tend to rotate $E$ modes generated by Thomson scattering into $B$ modes. Thus one would expect non-zero $BB$, $TB$ and $EB$ signals with power spectra similar to the corresponding $E$ mode spectra, but with a different amplitude. Since Faraday rotation is frequency dependent, one would expect the magnitude of this effect to depend on the frequency in which the CMB polarization is observed. 

The effects of such a large-scale magnetic field on the CMB power spectra were studied  \cite{scannapieco:1997} and \cite{scoccola:2004}. In \cite{scannapieco:1997} the authors looked at the $TB$ power spectrum and found that one could expect to put upper limits on a primordial magnetic field of the order $\B_0 \lesssim 10^{-8}$Gauss by the Planck experiment, where $\B_0$ denotes the strength of the magnetic field today. In \cite{scoccola:2004} the authors derived equations for all the $B$ power spectra using the total angular momentum method \cite{hu:1997}. Our method for calculating the different CMB power spectra is based on ref. \cite{scoccola:2004}, and we refer to this paper for further details on the notation and derivations. 

The effect of the Faraday rotation is conveniently described by the time independent parameter $F$, given by
\be{Fdef}
{}F =  0.7  \left( \frac{\B}{10^{-9}\textrm{Gauss}} \right) \left( \frac{10 \textrm{GHz}}{\nu} \right)^2,
\ee
where $\B$ denotes the strength of the magnetic field and $\nu$ is the frequency of the radiation. In \cite{scoccola:2004} the authors found that, if there are no $B$ modes in the absence of a magnetic field, the effect on the CMB $E$ and $B$ modes by including a non-zero $F$ can be expressed by
\bea
{}E_{lm} \pm iB_{lm} &=& \sum_{l',l''} \left[ \frac{(-1)^{l''-l}+1}{2} \right. \nonumber \\ 
&\pm& \left. \frac{(-1)^{l''-l}-1}{2} \right]  R(lm,l',l''m) \tilde{E}^{(l')}_{l''m}. 
\eea
where $R(lm,l',l''m)$ can be expressed in terms of Clebsch-Gordan coefficients:
\bea
{}R(lm,l',l''m) &=& i^{l'} \sqrt{\frac{2l''+1}{2l+1}} \nonumber \\
& \times & \langle l',l'';m-m'',m'' | l'l'';l,m \rangle \nonumber \\
& \times & \langle l',l'';0,2 | l'l'';l,2 \rangle. 
\eea
The effect of the magnetic field is hidden in the quantity $\tilde E^{(l')}_{lm}$: 
\bea
{}\tilde E_{lm}^{(l')} &=& -i^l \frac{4\pi}{2l+1} \int \frac{\ud^3 \mathbf{k}}{2 \pi^3}e^{i\mathbf k \cdot \mathbf x} Y_l^{m*}(\hat k) \sqrt 6 (2l+1) \nonumber \\
& \times & \int_0^{\eta_0} \ud \eta \dot \tau e^{-\tau}(2l'+1)j_{l'}(F\tau) P^{(0)}(\eta) \nonumber \\
& \times & \epsilon_l^{(0)} \left[ k(\eta_0 - \eta) \right]. \label{eq:Etilde}
\eea
This quantity only differs from the standard expression by the extra factor $(2l'+1)j_{l'}(F\tau)$, where $j_l$ denotes spherical Bessel functions of order $l$ and $\tau$ is the optical depth. Note that $\textrm{lim}_ {x \rightarrow 0} j_l(x) = 0$ for $l > 0$ such that all the introduced corrections reduce to 0 for $F=0$.  For exact definitions of the other quantities appearing in (\ref{eq:Etilde}) we refer to \cite{scoccola:2004}.

\begin{figure}[htbp]
\begin{center}
 \includegraphics[width=0.5\textwidth]{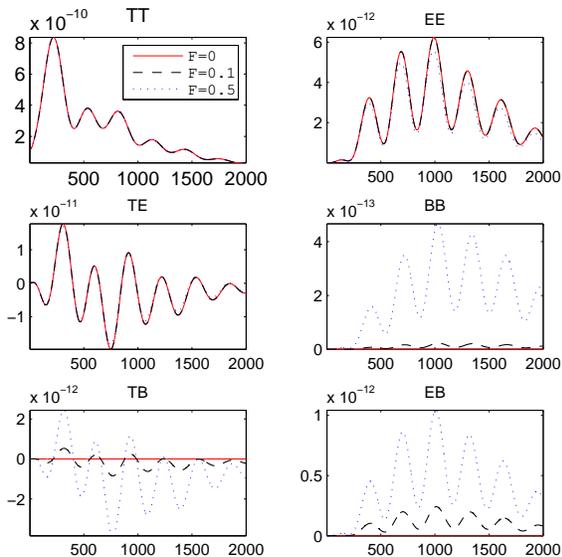} 
\caption{CMB power spectra for models with F={0, 0.1, 0.5}. The plotted quantity is $l(l+1)C^{XY}_l/2\pi$ as a function of multipole $l$. See text for definitions of the different $C^{XY}_l$.}
\label{fig:cls}
\end{center}
\end{figure}

With this definition of $\tilde E^{(l')}_{lm}$ the $TE$ power spectrum is now given by
\be
{}C_{l}^{TE} = C_l^{T \tilde E^{(0)}}.
\ee 
Resulting $TE$ power spectra for different values of $F$ are shown in Figure \ref{fig:cls}. 

For the other power spectra we will only study the limit $F \ll 1$, which is likely to be a good approximation. In this limit we can neglect contributions from $l' > 1$ in (\ref{eq:Etilde}). The $TB$ correlation can then be approximated by 
\be
{}C_l^{TB} = C_l^{T\tilde E^{(1)}} \frac{1}{2l+1} \sum_{m=-l}^{l} \left[ \frac{(l^2-m^2)(l^2-4)}{l^2(2l-1)(2l+1)} \right]^{1/2}
\label{eq:TB}
\ee
Here it should be stressed that the $TB$ correlations are non-diagonal. In the limit of $F \ll 1$ the contributions come from the $T_{l\pm1}B_l$ correlations. The $C_l^{TB}$ given in (\ref{eq:TB}) then represents a mean of the two off-diagonal contributions. Similarly, the $EB$ correlation can be expressed as
\be
{}C_l^{EB} = C_l^{\tilde E^{(0)} \tilde E^{(1)}} \frac{1}{2l+1} \sum_{m=-l}^{l} \left[ \frac{(l^2-m^2)(l^2-4)}{l^2(2l-1)(2l+1)} \right]^{1/2}
\label{eq:TE}
\ee
Also here the correlation is off-diagonal and eq. (\ref{eq:TE}) is the mean of the $E_{l\pm1} B_l$ contributions.

The autocorrelation polarization power spectra are given by
\bea
{}C_l^{EE} &=& C_l^{\tilde E^{(0)} \tilde E^{(0)}} \nonumber \\ 
{}C_l^{BB} &=& \frac 13 \left[  \frac{(l+1)^2-4}{(2l+3)(l+1)} C_{l+1}^{\tilde E^{(1)} \tilde E^{(1)}} +\frac{l^2-4}{l(2l+1)} C_{l-1}^{\tilde E^{(1)}\tilde E^{(1)}} \right]. \nonumber
\eea
The resulting power spectra have been calculated with a modified version of CAMB \cite{Lewis:1999bs, camb:2008}. In Figure \ref{fig:cls} the power spectra are plotted for different values of $F$ with the other cosmological parameters fixed to concordance $\Lambda$CDM values. From the figure it is obvious that both the $TB$, $EB$ and $BB$ power spectra are promising targets for constraining $\B_0$, especially since all these power spectra are expected to be zero in the case of $\B_0=0$.
Note that in the limit $F \ll1$  $C_l^{TB}$ and $C_l^{EB}$ scale linearly with $F$, such that the quantity $C_l^{TB}/F$ is constant. The amplitude of the $C_l^{BB}$ power spectrum will however scale like $F^2$ \cite{scoccola:2004}.

\section{Forecasts of Experimental constraints}
The aim of this work is to forecast the sensitivity of future CMB polarization experiments to the effect of a homogeneous magnetic field as described in Section \ref{sec:cls}. We will do this using a Fisher matrix approach. In our analysis we will assume all cosmological parameters except for $F$ to be known. This can be justified since our results are driven by the $TB$ and $EB$  correlations, which are zero when $F=0$. We will also make use of the $BB$ correlation,  but we note that the $BB$ power spectrum from a homogeneous magnetic field differs significantly from the $BB$ correlations from gravitational waves and lensing of $E$ modes.  The amplitude and shape of the different $B$  power spectra will indeed depend on the parameters governing the $E$ modes, but these parameters are already known with relatively good accuracy, and the precision will of course improve further with upcoming experiments. 

\subsection{The Fisher matrix approach}
Fisher matrices are a commonly used tool for forecasting experimental sensitivity.
Here one uses the curvature of the likelihood function around some fiducial model to predict the error of theoretical parameters given information on experimental parameters of the experiment being considered. In our case, since $F$  is the only parameter that is allowed to vary, the Fisher matrix will be $1 \times 1$, and the error in $F$ is given by
\be
{}\sigma_F^{-2} = \sum_l \frac{\left( \frac{\partial C_l^{XY}}{\partial F} \right)^2}{(\sigma_l^{XY})^2} \label{eq:sigmaF},
\ee
 where $X$ and $Y$ are $\{T, E, B\}$ and $\sigma_l^{XY}$ denotes the expected error in $C_l^{XY}$ in a given experiment. These errors are given by \cite{kamionkowski:1996, jaffe:2000}
 \bea
{}\sigma_l^{XY} &=&  \sqrt{\frac{1}{\fsky(2l+1)}}  \times \nonumber \\ 
	                   && \left[  \left( C_l^{XY}  +\fscan w_{XY}^{-1} B_l^{-2}  \right)^2 \right. \nonumber \\
                           &&+  \left(C_l^{XX}+\fscan w_{XX}^{-1} B_l^{-2}\right) \nonumber \\
                           &&\times \left. \left( C_l^{YY} + \fscan w_{YY}^{-1} B_l^{-2} \right) \right]^{1/2}. \label{eq:sigmas}
 \eea
 Here $\fsky$ denotes the fraction of the sky covered by the experiment.
This factor accounts for the smaller number of independent samples achieved when covering only parts of the sky. For a full-sky experiment like Planck, this factor is the part of the sky actually used in the analysis after the galaxy cut has been made.The fraction of the sky that is scanned by the experiment is denoted $\fscan$. It appears multiplying $w^{-1}$ because a smaller scanned sky fraction means longer integration time per pixel. Partial-sky experiments will of course aim to avoid the troublesome galaxy plane in their scanning strategy, making $\fsky \approx \fscan$ for such experiments.
 
A convenient notation to describe the noise in a pixel independent way is $w_{XX}^{-1} = \Omega_\textrm{pix}s^2/{t_\textrm{pix}} = (4\pi s^2)/(t_\textrm{pix}N_\textrm{pix}T_0^2)$ \cite{knox:1995,jaffe:2000}. Here $\Omega_\textrm{pix}$ is the angular pixel size, $t_\textrm{pix}$ is the time used observing each pixel and $N_\textrm{pix}$ is the total number of pixels. The detector sensitivity is denoted by $s$ and $T_0=2.73$K is the average sky temperature.  We have $w_{XY}=0$ for $X \neq Y$ \cite{kamionkowski:1996}. For $X = Y$ we have
 \be
{}w_{XX}^{-1} = 2.14 \times 10^{-15} t_\textrm{yr}^{-1} \left( \frac{s}{200 \mu \textrm{K}\sqrt\textrm{sec}}  \right)^2 \label{eq:winv},
 \ee
 where $t_\textrm{yr}$ is the observing time in years. From now on we denote $w_{T} \equiv w_{TT}$ and $w_P \equiv w_{EE} = w_{BB}$. For bolometric experiments  the polarization detectors will observe only the $Q$ or $U$ Stokes parameter at a time, leading to $w_P^{-1} = 2w_T^{-1}$.  

In eq. (\ref{eq:sigmas}) $B_l$ is the assumed Gaussian experimental beam, $B_l = e^{-l^2\sigma^2_\theta/2}$, where $\sigma_\theta$ is the beam width, $\sigma_\theta = \textrm{FWHM(beam)}/\sqrt{8 \ln 2}$. The $C_l$s appearing in (\ref{eq:sigmas}) are given in dimensionless $\Delta T/T$ units.
When estimating parameter errors from the $BB$ power spectrum, we use $\sigma_{F^2}^{-2}$ instead of $\sigma_F^{-2}$ in eqn. (\ref{eq:sigmaF}), since the $BB$ power spectrum scales as $F^2$. 

\subsection{Foregrounds}
In the procedure described above, we have neglected all effects from foregrounds. How much the foregrounds will affect the obtained limits on the magnetic field strength will depend on the specific experiments used, our understanding of the statistical properties of the different foregrounds, and whether a significant proportion of the foregrounds has been successfully subtracted from the CMB maps before the parameter analysis is performed. To do a  proper analysis of all the involved foreground effects is outside the scope of this work. However, we will use a simple foreground model to estimate the possible effects of foregrounds on our results. 

At the frequencies relevant here ($\nu \lesssim 100$GHz), the synchrotron emission from our galaxy is the dominant source of polarized foregrounds \cite{tegmark:1999}. For frequencies of order 100GHz the polarized emission from vibrating dust will also contribute. In the following analysis we will only include these two foreground components. We will then simply add the foreground signal to the CMB signal in eq. (\ref{eq:sigmas}) such that
\be
{}C_l^{XY} = C_{l,CMB}^{XY} + C_{l,sync}^{XY} + C_{l,dust}^{XY}, 
\ee
where  $C_{l,CMB}^{XY}$ is the CMB signal and $C_{l,sync}^{XY}$ and  $C_{l,dust}^{XY}$ come from the synchrotron and dust foregrounds, respectively. For both the foreground components we will assume a simple power law dependence \cite{tegmark:1999}:
\be
{}C^{XY}_{l,\{ \textrm{sync,dust} \}} = (pA)^2 l^{-\beta} \label{eq:foregrounds}
\ee
where $A$ is the amplitude of the unpolarized radiation and $p$ is the fractional amplitude of the polarized components and $\beta$ is a constant describing the scale-dependence. The numerical values of $A$, $\beta$ and $p$ for our dust and synchrotron foreground models are taken from \cite{tegmark:1999} and quoted in Table \ref{tab:foregrounds}.

\begin{table}[htb]
\begin{tabular}{lcccc}
\hline \hline
& & $TT$ & $TP$ & $PP$ \\
\hline
& $\beta$ {}& 2.4 & 1.9 & 1.4 \\
{Synchrotron} & $p$ & 1 & 0.3 & 0.13 \\ 
& $A$ & \multicolumn{3}{c}{101$\mu$K} \\
\hline
& $\beta$ & 3 & 1.95 & 1.3 \\  
{Dust} & $p$ & 1 & 0.0098 & 0.0024 \\
& $A$ & \multicolumn{3}{c}{24$\mu$K} \\
\hline
\end{tabular}
\caption{Numerical values for the foreground parameters in equation (\ref{eq:foregrounds}) for synchrotron and dust emission. Here $P=\{E,B\}$.}
\label{tab:foregrounds}
\end{table}

The foreground power spectra will also be frequency dependent.
For the synchrotron component, the frequency dependence is given by $\Theta_\textrm{sync}(\nu) \propto c(\nu) \nu^{-2.8}$, where $c(\nu)= (2 \sinh(x/2)/2)^2$ and $x=\nu/56.8$GHz.  $\Theta_\textrm{sync}(\nu)$ is normalized to be 1 for $\nu=19$GHz.

For the dust emission the frequency dependence is modelled by \cite{tegmark:1999}
\be
{}\Theta_\textrm{dust}=c(\nu)c_*(\nu)  \frac{\nu^{4.7}}{e^{h\nu/kT_\textrm{dust}}-1},
\ee
where $h$ is the Planck constant, $k$ is the Boltzmann constant, and $c_*(\nu) \propto 1/x^2$. The normalization is such that $\Theta_\textrm{dust}=1$ for $\nu=90$GHz.

%We will assume a perfect knowledge of the statistical properties of the synchro%tron foreground, that is, we know the exact shape of  $C_{l,sync}^{XY}$. This i%s of course an overly optimistic assumption. At the same ti
We are assuming that no foregrounds have been subtracted from the maps before the analysis is performed. This is a very pessimistic assumption, since the multi-frequency design of the different experiments will be used to subtract different foreground sources. Also, the upcoming C-BASS experiment \cite{cbass:2008} will make a 5GHz full-sky polarization map, tailored for synchrotron subtraction from CMB maps. The QUIJOTE experiment \cite{quijote:2008} will make measurements of microwave polarization on several low-frequency channels to further improve our knowledge on the polarized galactic foregrounds. We therefore believe that the actual ability of future CMB experiments to constrain primordial magnetic fields will be somewhere between the obtained results with and without the added foreground model. 

\subsection{Dependence on experimental parameters}
We will now turn our attention towards the different experimental parameters, and study how they affect the experimental sensitivity to $\B_0$. We  compare the limits obtainable from the $TB$, $EB$ and $BB$ power spectra.

The relevant experimental parameters that appear in $\sigma_l^{XY}$ are $\fsky, \fscan, w_T^{-1}, w_P^{-1} $ and $\sigma_\theta$.  When looking at a certain subset of the experimental parameters, we will fix the others to typical values for the 30GHz channel of the Planck experiment (see Table \ref{tab:experiments}). 

\begin{figure}[htbp]
\begin{center}
 \includegraphics[width=0.45\textwidth]{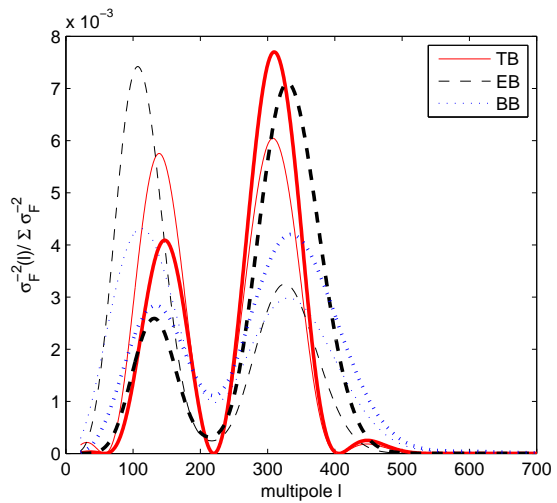} 
\caption{The contribution to $1/ \sigma_F^2$ from each multipole $l$ relative to the sum over all $l$s. The experimental parameters are set to Planck values. The red solid lines show the results for the $TB$ power spectrum, black dashed lines for the $EB$, and blue dotted lines for $BB$. The narrow lines show the results when no foreground is added, while the broad lines are the results when using the synchrotron and dust foreground model described in the text.}
\label{fig:lsigmas}
\end{center}
\end{figure}

\begin{figure}[htbp]
\begin{center}
 \includegraphics[width=0.45\textwidth]{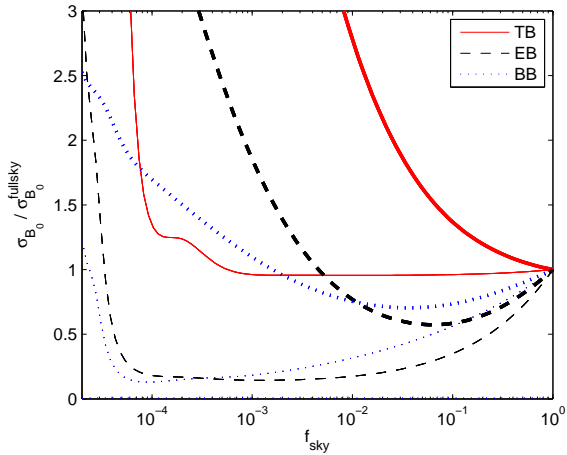} 
\caption{$\sigma_{\B_0}$ as a function of $\fsky$. Here the rest of the experimental parameters are set to typical Planck values. For each of the power spectra, $\sigma_{B_0}$ is normalized to 1  for $\fsky=\fscan=1$. The labeling of the graphs is the same as in Figure \ref{fig:lsigmas} }
\label{fig:fsky}
\end{center}
\end{figure}
 
The effect of changing $\fsky$ depends partly on which multipoles that contribute to the sum in eq. (\ref{eq:sigmaF}). In Figure \ref{fig:lsigmas} we have plotted $\sigma_F^{-2}$ for individual multipoles. It turns out that the main contribution to the sum comes from $l$s between 80 and 500. Comparing to the power spectra in Figure \ref{fig:cls}, this corresponds to the first peaks of the power spectra. Note that for the $BB$ power spectrum the distribution is more smeared out because of the $F^2$ dependence. We also see that the general pattern does not change much when adding the foreground model. From Figure \ref{fig:lsigmas} we would not expect that a large experiments multipole range would be important to place strong constraints on $\B_0$. 

In Figure \ref{fig:fsky} we illustrate the dependence of $\sigma_{\B_0}$ on $\fsky$ in a Planck-type experiment. Here we have set $\fsky = \fscan$ and only included multipoles $l > 180^o/\phi$ in the sum, where $\phi$ is the typical angular extension of an experiment with a given $\fsky$. We find significant changes in the behaviour with and without the foreground model added. For the foreground free model, the overall dependence on $\fsky$ is not very strong, as long as  $\fsky > 10^{-4}$. Also, we see that we get the best constraints from experiments with $\fsky \gtrsim 10^{-4}$, which corresponds to including $l \gtrsim 80$. This is exactly what we would expect from the $l$ dependence shown in Figure \ref{fig:lsigmas}. However, when the foreground is added, the importance of having a large $\fsky$ is more pronounced. This can be understood by looking at equation (\ref{eq:sigmas}). The gain of having a small $\fscan$ to obtain better sensitivity is less when the larger, foreground contaminated $C_l$s are added to the noise term. 

\begin{figure}[htbp]
\begin{center}
 \includegraphics[width=0.45\textwidth]{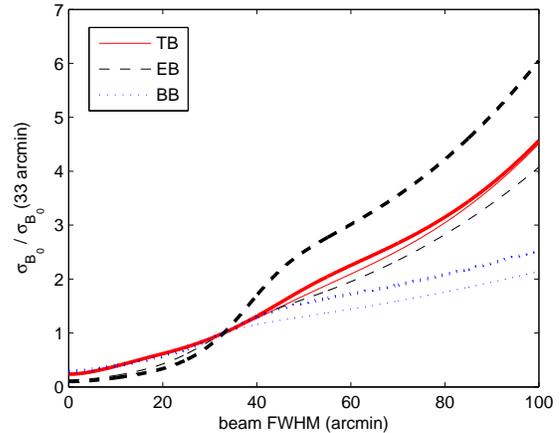} 
\caption{$\sigma_{\B_0}$ as a function of the beam FWHM. Here the rest of the experimental parameters are set to typical Planck values. $\sigma_{B_0}$ is normalized to 1 for the Planck beam of 33 arcminutes. The labeling of the graphs is the same as in Figure \ref{fig:lsigmas}}
\label{fig:beam}
\end{center}
\end{figure}

Next we have looked at the effect of changing the beam width for a Planck-type experiment. The dependence of $\sigma_{\B_0}$ on the beam FWHM is shown in Figure \ref{fig:beam}. We see that the dependence is approximately linear for beam widths between 0 and 100 arcminutes, and that $\sigma_{\B_0}$ changes  roughly an order of magnitude in this range for the $BB$ power spectrum, and slightly more when using the $TB$ and $EB$ power spectra. There is no big change in the importance of the beam size when adding the foreground model, but we notice that there is a slightly larger relative gain of having a smaller beam when including foregrounds.  

\begin{figure}[htbp]
\begin{center}
 \includegraphics[width=0.45\textwidth]{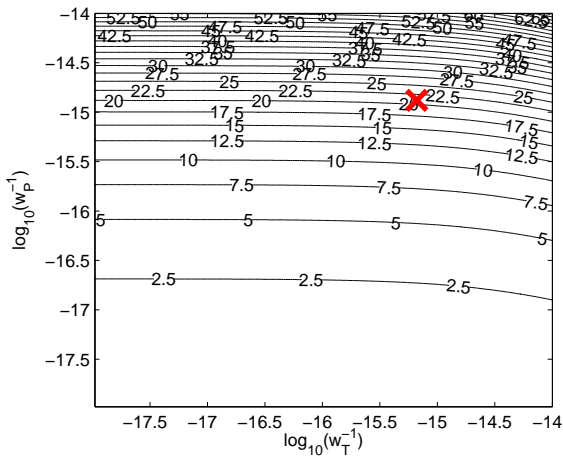} 
\caption{Contours of $\sigma_{\B_0} /10^{-10}$Gauss  in the $w_P - w_T$ plane when using the $TB$ power spectrum. Here the rest of the experimental parameters are set to typical Planck values. The Planck values of $w_P$ and $w_T$ are marked with a red cross. No foregrounds are added here.}
\label{fig:ws}
\end{center}
\end{figure}

\begin{figure}[htbp]
\begin{center}
 \includegraphics[width=0.45\textwidth]{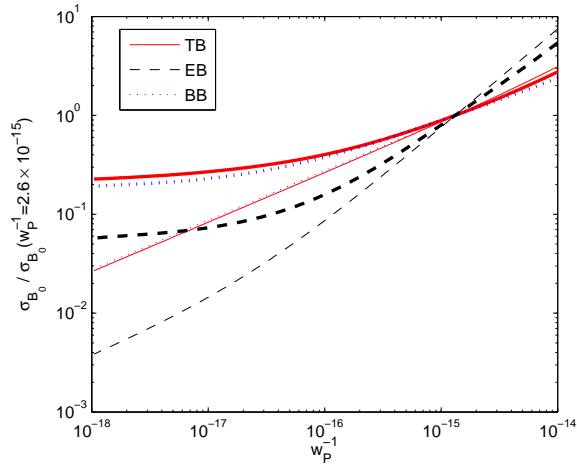} 
\caption{$\sigma_{\B_0}$ as a function of $w_P^{-1}$ for the $TB$, $EB$ and $BB$ power spectra. Here $\sigma_{B_0}$ is normalized to 1 for $w_P = 13.2\times 10^{-16} $ (the Planck value). Other experimental parameters are set to Planck values. For the $TB$ spectrum we have used  $w_T^{-1} = 0.5 w_P^{-1}$.  The labeling of the graphs is the same as in Figure \ref{fig:lsigmas}}
\label{fig:wp}
\end{center}
\end{figure}

Large effects are also achieved when varying $w_P^{-1}$ and $w_T^{-1}$. In Figure \ref{fig:ws} we show contours of $\sigma_{\B_0}$ in the $w_T^{-1}-w_P^{-1}$ plane when using the $TB$ power spectrum. Here no foregrounds are added. It is obvious that a good experimental sensitivity and/or a long integration time is crucial to place tight constraints on $\B_0$. From the plot we see that the main importance is to have good polarization sensitivity, but as already mentioned, $w_P^{-1}$ and $w_T^{-1}$ are proportional in standard bolometer experiments. In Figure \ref{fig:wp} we show the dependence of $\sigma_{B_0}$ on $w_P^{-1}$ when using the different power spectra ($w_T^{-1} = 0.5w_P^{-1}$ in the $TB$ case). We see that improving $w_P^{-1}$ is more important for $EB$ than in the case of $TB$ and $BB$.  Since the $BB$ power spectrum scales as $F^2$, the potential for using it to constrain $\B_0$ will be less important as the sensitivity improves. Another important feature in Figure \ref{fig:wp} is the large difference between the behavior with and without foregrounds included.  We observe a significantly smaller gain in sensitivity to $\sigma_{\B_0}$ by decreasing $w_P^{-1}$ when the foreground model is included. 

\begin{figure}[htbp]
\begin{center}
 \includegraphics[width=0.45\textwidth]{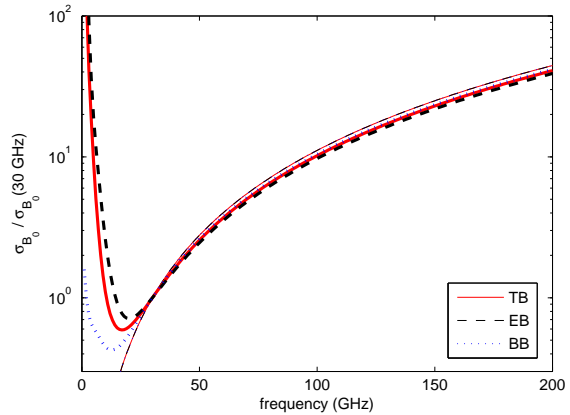} 
\caption{Here $\sigma_{\B_0}$ is plotted as a function of the observed frequency. The labels are the same as in Figure \ref{fig:lsigmas}. The values are normalized to 1 for 30GHz.}
\label{fig:freq}
\end{center}
\end{figure}

In Figure \ref{fig:freq} we have plotted $\sigma_{\B_0}$ as a function of the observed frequency. In the foreground free scenario, this will simply result in a $\nu^{-2}$ dependence. Including the synchrotron and dust foregrounds makes the picture more interesting, as the amplitude of the synchrotron power spectrum has an even stronger negative correlation with the frequency than the Faraday rotation effect.  We see that there is a
real competition between different effects at low frequencies. 
For an Planck like experiment, observing at frequencies lower than 20GHz will in fact be less sensitive to $\B_0$ with our foreground model.

To summarize, we need experiments characterized by good polarization sensitivity and preferably also a small beam width and low frequency to be able to place tight constraints on $\B_0$. With unsubtracted foregrounds, the importance of a small beam persist, while the importance of a small $w_P^{-1}$ decreases. Also, with foregrounds included, there is preference for a larger observed sky fraction than in the foreground free scenario. 

\subsection{The experiments}

We now consider some specific upcoming CMB experiments and forecast their ability to constrain $\B_0$. We will concentrate on the lowest frequency channels of the experiment, given that the effect of $\B_0$ carries such a strong frequency dependence. 

The Planck experiment \cite{planck:2005} is a satellite based full-sky experiment to be launched in 2008. Some of the experimental characteristics are listed in Table \ref{tab:experiments}. In addition we will assume $\fsky=0.8$ due to the galaxy cut that has to be made before the data analysis. We have considered a 14 months survey.  For Planck we will use $l$s between 2 and 2500 in our analysis.

The QUIET experiment \cite{quiet:2008, samtleben:2008} is a ground based experiment dedicated to CMB polarization, which will start taking data in 2008. Instead of using bolometers, QUIET uses arrays of HEMT based receivers that measure $Q$ and $U$ simultaneaously. The experiment is divided into two phases. The first phase (P1) will last for 2 years using a 1m telescope. The following phase 2 (P2) will use a set of three 2m telescopes and one 7m telescope. Experimental parameters \cite{lawrence:2004} are listed in Table \ref{tab:experiments}. For QUIET we will use $l$s between 50 and 2500. 

The Clover experiment \cite{taylor:2006, north:2007} is another ground based CMB polarization experiment. The 96GHz telescope which we will refer to here, will start taking data in 2009, and collect data with an effective integration time of $\sim 0.8$ years.  For Clover we use $l$s between 20 and 1000. 

We will also look at two balloon-borne experiments; Spider \cite{mactavish:2007} and EBEX \cite{oxley:2005}. Spider will cover about 60\% of the sky during $\sim25$ flight-nights, starting from Australia  in December 2009. We apply multipoles between $l=4$ and $l=500$ for the Spider experiment. We will assume $\fsky=0.5$  For other experimental parameters we refer to Table \ref{tab:experiments}. 
EBEX will having a total integration time of about 14 days, and start operating in 2008. It will cover multipoles between 20 and 1500, operating on three frequency bands, the lowest one at 150GHz. Further experimental characteristics \cite{johnson:2008} are given in Table \ref{tab:experiments}.

The ground based QUIJOTE experiment \cite{quijote:2008} is designed to determine the characteristics of the polarized galactic foregrounds, covering about 10\% of the sky on the northern hemisphere. Unlike the C-BASS experiment, QUIJOTE is sensitive to frequencies as high as 30GHz, which should make it suitable for detecting the polarization signal from primordial magnetic fields. This is provided that it surveys patches sufficiently far away from the highly foreground dominated galactic plane. Experimental characteristics \cite{genova:2008} for the QUIJOTE 30GHz channel are given in Table \ref{tab:experiments}.

\begin{table}[htb]
\begin{tabular}{lccccc}
\hline \hline
Experiment & $\nu$(GHz) &$\fscan$& beam(') & $\frac{w_T^{-1}}{10^{-16}}$ & $\frac{w_P^{-1}}{10^{-16}}$\\
\hline
Planck             &30 & 1       & 33 & 6.6  & 13.2 \\
 QUIET P1       &40 &  0.03 & 41 & 8.1  & 0.98 \\
QUIET P2 2m &40 & 0.04 & 23 & 0.67 & 0.062 \\
QUIET P2 7m &40 & 0.01 & 9   & 1.34 &  0.12 \\
Clover              &97 & 0.23	& 8   &  0.045 & 0.090 \\
Spider             & 96 & 0.6   & 58  & 0.12 & 0.24 \\   
EBEX              &150 & 0.01 & 8    &  0.18 & 0.35 \\
QUIJOTE        &30  &  0.1   &  22 & 1.3  & 2.7 \\
\hline
\end{tabular}
\caption{Experimental characteristics for some CMB polarization experiments. The beam size refers to the FWHM value in arcminutes.}
\label{tab:experiments}
\end{table}

\section{Results}
Throughout our analysis we have used as a fiducial cosmological model a standard flat $\Lambda$CDM model based on the mean parameter values for the WMAP 5 year results \cite{dunkley:2008} and defined by $\{\Omega_bh^2, \Omega_{c}h^2, h, \tau, n_s, A_s\} = \{ 0.023, 0.11, 0.72, 0.09, 0.96, 2.3\times10^{-9}\}$. 
%\{ 0.022, 0.11, 0.73, 0.09, 0.96, 2.3\times10^{-9}\}$ 
The parameter definitions used correspond to the standard definitions in CAMB. Here $\Omega_b h^2$ and $\Omega_c h^2$ are the physical densities of baryons and cold dark matter, respectively (we assume massless neutrinos). The Hubble parameter is parametrized by $H_0 = 100 h \textrm{km s}^{-1}\textrm{Mpc}$, and $\tau$ is the optical depth at reionization. The primordial power spectrum is parameterized by the amplitude $A_s$ (at 0.05 Mpc$^{-1}$) and the tilt $n_s$. We do not expect our results to change significantly by minor changes in this fiducial model. 

The resulting forecasts for $\sigma_{\B_0}$ for the different experiments are summarized in Table \ref{tab:results}.
\begin{table}[htb]
\begin{tabular}{lccc}
\hline \hline
Experiment    & \quad $TB$  \quad & \quad  $EB$  \quad & $BB$ \\
\hline
Planck              &  2.1(2.5)     &  8(12)       &  15(18)     \\
QUIET P1        &   1.1(3.6)   &     0.7(3.1)   &  3.3(10)     \\
QUIET P2 2m  &  0.15(1.2)   &  0.07(0.8)  &   0.5(4.3)  \\
QUIET P2 7m \quad &  0.10(1.0)    &  0.041(0.7)   &  0.31(3.7) \\
Clover              & 0.6(0.7)      & 0.22(0.26)  & 3.3(3.9) \\
Spider             &   6(6)        &     3.6(4.5)     &   25(29)       \\
EBEX              &   2.4(3.9)   & 0.9(1.6)    & 7(13)               \\
QUIJOTE        &  0.5(1.5)    &  0.34(2.0) & 2.4(6)     \\
\hline
\end{tabular}
\caption{Expected values of $\sigma_{\B_0}/10^{-9}$Gauss  for the different experiments based on the experimental characteristics from Table \ref{tab:experiments}. The numbers in parenthesis corresponds to the results when including the foreground model, while results from the foreground free analysis are quoted outside the parentheses.}
\label{tab:results}
\end{table}

Let us first concentrate on the results obtained without the foreground model. 
We can identify which power spectrum gives the best results depends on the experiment under consideration. That $TB$ is less important for QUIET than for e.g. Planck is easily understood by the fact that QUIET has much better $w_P^{-1}$ relative to $w_T^{-1}$ than the bolometer based experiments. But we have also seen that the results from the $EB$ and $BB$ power spectra depend more strongly on $w_P^{-1}$ and $w_T^{-1}$ than does the $TB$ power spectrum. This explains e.g. why we for Planck find the best limits from using $TB$, while the best limits come from $EB$ for Clover. The most impressive results come from the QUIET experiment's phase 2 telescopes. Here the EB power spectrum can constrain $\B_0$ by $\sigma_{B_0} = 7\times10^{-11}$ and  $\sigma_{B_0} = 4 \times10^{-11}$ from the 2m and 7m telescopes, respectively. This is approximately a factor 50 better than what we find for Planck $TB$ and Spider, a factor 20 better than EBEX and about a factor 5 better than Clover and QUIJOTE.

Adding the foreground model, this picture changes significantly. The results from QUIET P2 now weakens by more than an order of magnitude, while the changes for the other experiments are typically by  $\sim 10\%$. The reason why QUIET is much more sensitive to this foreground is caused by several factors. Firstly, the QUIET experiment covers a small fraction of the sky, and we have seen that while this can be beneficial in a foreground-free scenario, the opposite is the case when the foreground is present. Also, QUIET operates at a relative low frequency compared to for example Clover and EBEX, which makes the synchrotron foreground much stronger, as its amplitude has a $\nu^{-2.8}$ dependence. We have also seen that the effect of having good experimental sensitivity is less important in the presence of foregrounds. The result is that when including the foreground model, we find the best results from Clover $EB$, giving  $\sigma_{B_0} = 2.6\times10^{-10}$Gauss, which is a factor 3 better than what we find for QUIET, a factor 6 better than EBEX and an order of magnitude better than Planck.  

Which of the results with or without the foreground are more close to the real performance of the experiments depends on how well the foreground can be subtracted from the maps. This will be of special importance for QUIET and less important for the other experiments considered here. Also, we have assumed perfect knowledge of the statistical properties of the foreground model. When analyzing a real experiment, these uncertainties must be taken into account and this may alter the results, especially for a "foreground sensitive" experiment like QUIET. We also note that the results in Table \ref{tab:results} are quoted for each individual power spectrum. Simply combining the results from the different power spectra (and experiments) will of course improve the limits slightly. 
 
\section{Conclusions}

In this work we have calculated CMB power spectra for a universe with a primordial, homogeneous magnetic field. We introduced a simple foreground model, and applied a Fisher matrix approach to explore how the experimental sensitivity to such a magnetic field depended on different experimental parameters. We then attempted to forecast our ability to constrain $\sigma_{B_0}$ for various future CMB polarization experiments.

In our foreground model we have assumed perfect knowledge of the statistical behavior of the foregrounds, but that none of it has been removed from the CMB maps prior to the parameter analysis. We argue that this can be regarded as a conservative approach. Including this foreground model we find that among the experiments considered here, Clover has the best prospects of constraining $\B_0$, and we find $\sigma_{\B_0} \approx 3 \times 10^{-10}$Gauss in this case. When assuming that all foregrounds are perfectly subtracted from the maps, we find QUIET to be the most promising experiment, giving $\sigma_{\B_0} \approx 4 \times 10^{-11}$Gauss. This is two orders of magnitude better than the present upper limits on $\B_0$ from using the homogeneous anisotropy of CMB \cite{barrow:1997, giovannini:2005}. 
For the QUIET experiment, the results rely heavily on the ability to successfully subtract the foregrounds, while the other experiments under consideration are less dependent on this. 

To conclude, our study shows that we will be able to significantly improve our knowledge of possible primordial homogenous magnetic fields in the coming years. A positive detection will impact both the construction of inflationary models and our understanding of the origins of magnetic fields in galaxy clusters.

\acknowledgments
We thank Ruth Durrer for discussions,
Bradley R. Johnson for providing details on the EBEX experiment, and Ricardo T. G{\'e}nova-Santos for details on the QUIJOTE experiment. 
JRK is grateful to the University of Oxford Astrophysics for their hospitality during the work on this project. JRK acknowledges financial support from the Research Council of Norway.

%%%%%%%%%%%%%%%%%%%%%%%%%%%%%%%%%

%\bibliography{cites}

\end{document}